\documentclass[amsmath,amssymb,superscriptaddress,aps,a4paper,prl,reprint]{revtex4-1}

\usepackage[dvips]{color}
\usepackage{graphicx} 
\usepackage{amsmath}
\usepackage{amssymb}
\usepackage{amsfonts}
\usepackage{lmodern}
\usepackage{textcomp}
\usepackage[T1]{fontenc}
\usepackage{gensymb} 
\usepackage[seperr, load={}]{siunitx}

\newcommand{\twobytwo}[4]{\left( \begin{array}{cc} #1 & #2 \\ #3 & #4 \end{array}\right)}
\newcommand{\phiS}{\tilde{\varphi}_S}
\newcommand{\phig}{\gamma}
\newcommand{\phid}{\xi}
\newcommand{\ket}[1]{|#1\rangle}

\newcommand{\adiabstates}{\ket{g},\ket{e}}
\newcommand{\Article}{Letter}

\begin{document}

\title{Geometric Landau-Zener interferometry}

\author{S. Gasparinetti}
	\email{simone@boojum.hut.fi}
	\affiliation{Low Temperature Laboratory, Aalto University, P.O. Box 15100, FI-00076 Aalto, Finland}
\author{P. Solinas}
	\affiliation{Low Temperature Laboratory, Aalto University, P.O. Box 15100, FI-00076 Aalto, Finland}
	\affiliation{Department of Applied Physics/COMP, Aalto University, P.O. Box 14100, FI-00076 AALTO, Finland}
\author{J. P. Pekola}
	\affiliation{Low Temperature Laboratory, Aalto University, P.O. Box 15100, FI-00076 Aalto, Finland}
\date{\today}

\begin{abstract}
We propose new type of interferometry, based on geometric phases accumulated by a periodically driven two-level system undergoing multiple Landau-Zener transitions. As a specific example, we study its implementation in a superconducting charge pump. We find that interference patterns appear as a function of the pumping frequency and the phase bias, and clearly manifest themselves in the pumped charge. We also show that the effects described should persist in the presence of realistic decoherence.
\end{abstract}

\maketitle

A driven quantum two-level system traversing an avoided energy-level crossing can undergo nonadiabatic transitions, known as the Landau-Zener effect.
If more than one crossing is involved and the dynamics is overall coherent, then transition paths can interfere according to the different phase accumulated by the ground and excited-state wavefunctions between subsequent crossings. This phenomenon, sometimes referred to as Landau-Zener-St\"uckelberg (LZS) interferometry \cite{Shevchenko2010}, was first observed in atomic and optical systems, and recently proposed \cite{Shytov2003} and measured also in superconducting qubit systems \cite{Oliver2005,Sillanpaa2006,Wilson2007,Izmalkov2008,Sun2011}.
In all these realizations, the system is driven in such a way that the interference effects have a purely dynamical nature. In general, though, a quantum state subject to steered evolution acquires both a dynamic and a geometric phase. While the study of geometric phases in solid-state systems is an active field of research \cite{Falci2000,Leek2007,Mottonen2008}, their relevance to LZS interferometry has so far been unexplored \cite{Bow_note}.

In this \Article, we elucidate the link between LZS interference and geometric phases, opening new possibilities for the geometric control of quantum systems. Our results apply to a broad range of devices, namely those for which the parametric driving possesses a nontrivial geometric structure and the induced energy-gap modulation presents multiple avoided crossings.
As a pertinent example, we consider a superconducting charge pump, the Cooper-pair sluice \cite{Niskanen2003}.
The connection between Cooper-pair pumping and geometric phases was highlighted in previous theoretical works both for the abelian \cite{Aunola2003, Mottonen2006} and nonabelian \cite{Brosco2008} case, yet always in the adiabatic limit, where the system stays in the instantaneous ground state and excitations are treated as small corrections \cite{Russomanno_note}. We instead consider higher frequency regimes and predict the appearence of interference patterns depending on the pumping frequency and the superconducting phase bias, the latter embodying the geometric contribution to interference. We then show that LZS resonances directly manifest themselves in the pumped charge, which is an advantage of using a charge pump rather than a conventional qubit as an interferometer. Finally, we introduce decoherence in our model and show that interference effects are still detectable. This should make our proposal feasible for experimental observation.

\begin{figure}
\center
\includegraphics{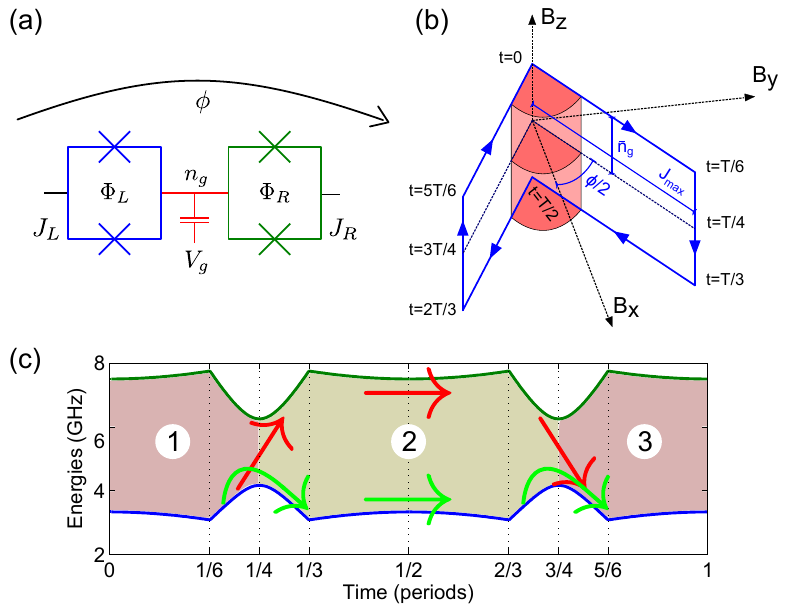}
\caption{(Color online) (a) Schematic drawing of the ``sluice''. (b) Effective magnetic field corresponding to the pumping cycle considered in this \Article.
(c) Adiabatic (instantaneous) energy versus time.
Avoided level crossings occur at times $t_1=T/4$ and $t_2=3T/4$. Green and red arrows outline two possibly interfering paths.}
\label{fig:1}
\end{figure}

The Cooper-pair sluice, schematically shown in Fig.~1(a), consists of a superconducting island coupled to the leads by two superconducting quantum interference devices (SQUIDs), whose Josephson energies $J_{L,R}$, can be tuned by changing the magnetic fluxes $\Phi_{L,R}$. A gate electrode capacitively coupled to the island is used to induce a polarization charge $n_g$ on the latter, thereby providing a third control parameter. During a pumping cycle, the parameters are steered so as to couple the island to the left lead, attract a Cooper-pair, switch the coupling to the right lead, and release the Cooper-pair \cite{cycle_note}.
We will assume that the superconducting phase difference $\phi$ across the device is kept constant. This can be achieved by shunting the sluice with a large Josephson junction. In this case, the switching statistics of the additional junction also provides a way to measure the pumped charge \cite{phi_note}.

The device is operated in the Coulomb-blockade regime $E_C \gg J_{max}$, where
$J_{max}= \max\{J_L, J_R\}$ and $E_C$ is the charging energy of the island.
The system dynamics is then best described in the basis of eigenstates of charge on the island. Also, as long as $n_g$ stays close to the degeneracy point $1/2$, only two such states are relevant, namely those with zero and one excess Cooper pair on the island.
This allows us to use a pseudo-spin formalism and write the sluice Hamiltonian as $H=\vec{\sigma}\cdot \vec{B}$, where $\{\sigma_i\}$ are the Pauli matrices and the effective magnetic field $\vec{B}$ has components
\begin{eqnarray}
 B_x(t)&=& J_+ (t)\cos\frac{\phi}{2}\ , \\
 B_y(t)&=& J_-(t) \sin\frac{\phi}{2}\ , \\ 
 B_z(t)&=& E_C \left[1/2 -n_g(t) \right]\ ,
\end{eqnarray}
where we put $J_\pm (t) = J_L (t) \pm J_R(t) $.
As $\vec{B}$ is steered along the path shown in Fig. 1(b), it spans a solid angle which is responsible for the geometric effects under discussion. This situation is clearly different from that considered in e.g. Refs. \cite{Oliver2005, Sillanpaa2006}, where $\vec{B}$ moves on a definite plane (say, $x$-$z$), leaving no room for non-trivial geometric effects to take place.


In Fig.~1(c) we plot the energies of the adiabatic states $\adiabstates$ as a function of time for a pumping cycle, obtained by instantaneous diagonalization of the Hamiltonian. The avoided level crossings at $t=T/4,3T/4$ ($T$ is the pumping period) correspond to the gate charge crossing the degeneracy point.
The probability of a nonadiabatic (Landau-Zener) transition at such a crossing is given by
$P_{LZ}=e^{-2\pi\delta}$, where the adiabatic parameter $\delta$ depends on the velocity at which the crossing is traversed and on the energy gap at the crossing \cite{Shevchenko2010}. For our case, $\delta=\pi J_{max}^2/ \left(48 E_C \bar{n}_g h\nu \right)$, where $\bar{n}_g=\max \{ n_g \}-1/2$ and $\nu=1/T$.

In the limits $E_C \gg J_{max}$ and $h\nu \lesssim E_C\bar{n}_g$, nonadiabatic transitions are strongly localized at level crossings.
The system dynamics can thus be seen as a sequence of adiabatic evolutions and localized transitions.
For this reason, the calculation is most conveniently performed in the adiabatic basis $\{\adiabstates\}$.
In the so-called adiabatic-impulse model \cite{Damski2006,Shevchenko2010}, Landau-Zener tunneling at anticrossings is treated as instantaneous and described in the adiabatic basis by a transfer matrix of the form:
\begin{equation}\label{eq:LZmat}
N_{LZ} =\twobytwo{\sqrt{1-P_{LZ}} e^{i\phiS}}{- \sqrt{P_{LZ}}}{\sqrt{P_{LZ}}}{\sqrt{1-P_{LZ}} e^{-i\phiS}}\ ,
\end{equation}
where $\phiS= \delta (\log\delta -1 ) + \arg \Gamma (1-i\delta) - \pi/4$ is the impulsive phase acquired by the adiabatic states in traversing the crossing ($\Gamma$ is the gamma function) \cite{Kayanuma1997,Sillanpaa2006}.

For each adiabatic segment $j=1,2,3$ in Fig.~\ref{fig:1}(c), evolution from time $t_{j-1}$ to $t_j$ is described by a diagonal matrix of the form:
$U_j=\exp\left[i \varphi_j \sigma_z\right]\ ,$
where $\varphi_j$ is the total phase difference acquired by the adiabatic states. The latter can be written as
$\varphi_j = \phid_j +\phig_j$, where we have distinguished a dynamic ($\phid_j$) and a geometric ($\phig_j$) contribution. The dynamic phase difference $\phid_j$ is given by:
\begin{equation}\label{eq:phid}
\phid_j=\frac{1}{2\hbar} \int_{t_{j-1}}^{t_j} dt \sqrt{|H_{11}-H_{22}|^2 + 4|H_{12}|^2}\ .
\end{equation}

By contrast, the gauge-invariant, noncyclic geometric phase difference $\phig_j$ can be calculated as \cite{GarcadePolavieja1998}:
\begin{equation}\label{eq:phig}
\phig_j = \frac{i}{2}\int_{t_{j-1}}^{t_j} dt \left[ \left \langle g \left| \frac{d}{dt} \right| g \right \rangle -\left \langle e \left| \frac{d}{dt} \right| e \right \rangle \right] \ .
\end{equation}
This phase is uniquely determined by the path drawn by the system in parameter space, and reduces to the Berry phase for cyclic ground-state evolution \cite{Berry_note}.

Putting things together, the evolution operator over a period can be calculated as
\begin{equation}\label{eq:Utot}
U = U_3N_{LZ}U_2N_{LZ}U_1= U_3\twobytwo{\alpha}{-\beta^*}{\beta}{\alpha^*} U_1 \ ,
\end{equation}
where
\begin{eqnarray}
\alpha & = &  \left[ (1-P_{LZ}) e^{2i\phiS+i\phid_2+i\phig_2}-P_{LZ}e^{-i\phid_2-i\phig_2} \right]  \ , \\
\beta & = &2 \sqrt{P_{LZ}(1-P_{LZ})} \cos\left(\phiS+\phid_2+\phig_2\right) \ .
\end{eqnarray}
$U_1$ and $U_3$ play no role in the upcoming resonance condition and will not be considered further.
From \eqref{eq:Utot}, we can calculate the excitation probability after one period starting from the ground state. This is given by:
\begin{equation}\label{eq:rescond}
P=4P_{LZ}(1-P_{LZ}) \cos^2\left(\phiS+\phid_2+\phig_2\right)\ .
\end{equation}
This probability oscillates between $0$ and $4P_{LZ}(1-P_{LZ})$ as a function of the accumulated phase. In the fast-passage limit, $\delta \ll 1$, we can approximate $\phiS \approx -\pi/4$ (in the adiabatic limit $\delta \rightarrow \infty$ and $\phiS \rightarrow -\pi/2$). 

We now make our discussion specific by considering the pumping cycle of Fig.~1(b)
\cite{expression_note}. Up to the first order in $[J_{max}/(E_C\bar{n}_g)]^2$, we find:
\begin{eqnarray}
\phid_2 & = & \frac{5\pi}{6} \frac{E_C\bar{n}_g}{h\nu}\ , \\
\phig_2 & = & \phi/2 \ .
\end{eqnarray}

As expected, the dynamic phase $\xi_2$ is inversely proportional to the pumping frequency $\nu$. By contrast, the geometric contribution $\gamma_2$ does not depend on $\nu$, and in this particular case equals half the superconducting phase bias $\phi$.
We have thus derived a resonance condition involving the superconducting phase bias $\phi$ and the pumping frequency $\nu$. In particular, in the region where $\phiS$ attains a constant value, the resonances drift in the $\phi-\nu$ plane as branches of hyperbolae.

This analysis predicts the position of resonances and explains their origin. Its regime of validity lies in between the strictly adiabatic and the fully nonadiabatic one. As a matter of fact, a lower bound for the pumping frequency is set by the requirement for time evolution to be coherent over one pumping cycle. On the other hand, at frequencies comparable to the adiabatic level spacing ($h\nu \approx E_C \bar{n}_g$) transitions are no longer restricted to the degeneracy points and the adiabatic-impulse model is expected to break down.

The superconducting phase bias $\phi$ enters the resonance condition through the geometric phase accumulated between subsequent transitions. This relationship is trivial for the case considered, as the geometric phase is simply proportional to $\phi$. Yet, this example clearly illustrates the role of geometric phases in Landau-Zener interference.
In particular, by choosing the pumping frequency so that $\phiS+\phid_2$ is an integer multiple of $\pi$, the dynamic contribution in \eqref{eq:rescond} is washed out, resulting in a purely geometric Landau-Zener interference effect.

\begin{figure}
\center
\includegraphics{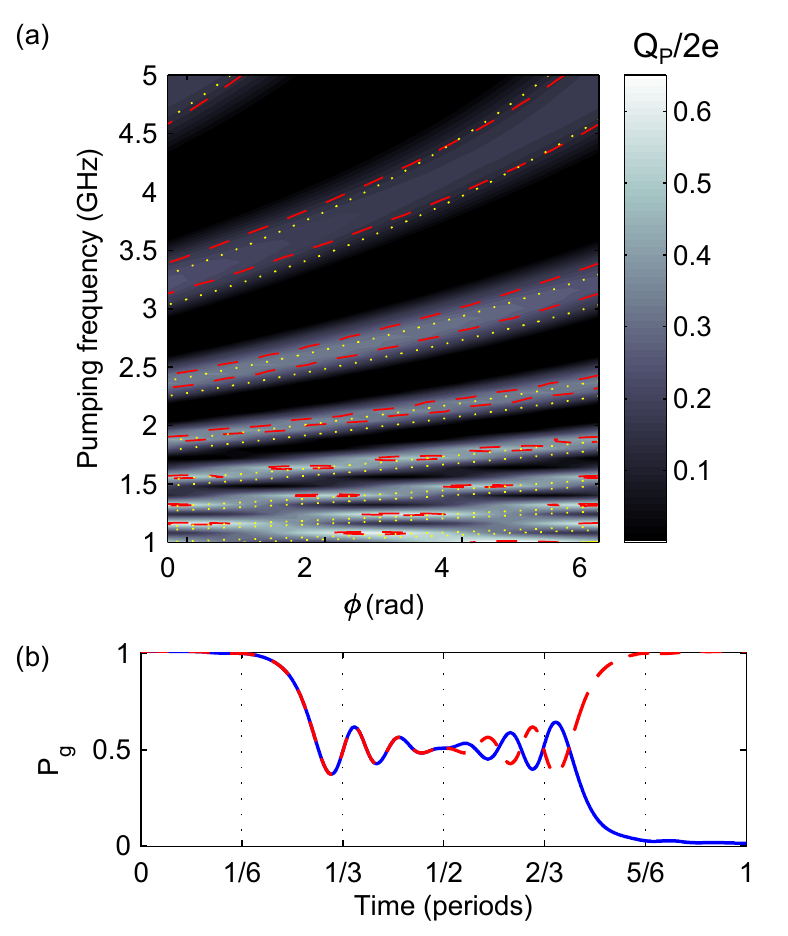}
\caption{(Color online) (a) Pumped charge (in units of $e$) after one cycle versus phase bias and frequency. The parameters are: $E_C/k_B=2.5$~K, $J_{max}=0.1 E_C$, $0.3 \leq n_g \leq 0.7$.
Dashed lines enclose the regions where the ground state population at the end of the cycle is at least 0.9.
Dotted lines have the same meaning but they are calculated according to \eqref{eq:rescond} and in the fast-passage approximation $\phiS\approx -\pi/4$. (b) Ground-state population versus time for a case of destructive (dashed line) and constructive (solid line) interference. The pumping frequency is 1.56 GHz for both cases, the phase biases are 0.22 and 3.36, respectively.}
\label{fig:2}
\end{figure}

We now proceed to show that the predicted resonances manifest themselves in the charge pumped by the device, thus providing the most straightforward way of observing them. To do so, we first obtain the full system dynamics from numerical solution of the Schr\"odinger equation.
We then calculate the pumped charge by integrating the instantaneous current operator \cite{Qp_note}.
In Fig. 2 (a), we plot the pumped charge over a period versus the phase bias $\phi$ and the pumping frequency $\nu$.
The parameters are chosen so as to be consistent with our model. In particular, microscopic excitations in the superconducting circuit can be neglected provided $h f_{eff}\ll \Delta$, where $f_{eff}$ is the effective frequency of the driving fields and $\Delta$ the superconducting gap. Furthermore, for small values of $\bar{n}_g$ the system is sufficiently anharmonic for the two-level approximation to hold in the given frequency range.
The lines drawn on top of the image plot correspond to 90\% probability of the system being in the ground state at the end of the cycle. Dashed lines are calculated numerically, dotted lines according to \eqref{eq:rescond}. The strong correlation between the ground state population and the pumped charge demonstrates the possibility to access interference patterns simply by measuring the latter. Moreover, the accuracy of the approximations made in deriving \eqref{eq:rescond} is confirmed by the good agreement between analytical and numerical calculations.

In Fig. 2 (b), we show the time evolution of ground-state populations for one case of constructive and one of destructive interference.
In both cases there is a population transfer to the excited state after the first crossing. Yet, while constructive interference (solid line) enhances the excitation after the second crossing, destructive interference (dashed line) brings the system back to the ground state. In particular, this implies that for a given pumping frequency, the phase bias can be chosen so as to pump a significant fraction of Cooper pairs (about 0.5 in this case) even in the nonadiabatic regime.

A complementary and instructive way to understand these features is provided by Floquet analysis \cite{Grifoni1998}. In fact, we can explicitly calculate the quasienergy spectrum by diagonalizing the evolution operator $U$ in \eqref{eq:Utot}. We find that destructive resonances occur at exact quasienergy crossings, where time evolution over a period is trivial and tunneling between adiabatic states is dynamically frozen. This phenomenon is known as coherent destruction of tunneling \cite{Grossmann1991}. At the opposite end, constructive interference enhances such transitions, resulting in Floquet states being the maximal mix of the adiabatic ones. This is revealed in the quasienergies as the opening of a gap, similarly to a time-independent system with a coupling interaction switched on.

The LZS interferometry discussed above is a unitary and coherent process. However, in any experimental implementation the undesired and unavoidable coupling to external degrees of freedom leads to decoherence and thus it affects the observed pumped charge. This fact precludes a measurement of coherent LZS pumping over a great number of cycles and must be taken into account in any experimental proposal.
To this end, we will now discuss the case in which the system is affected by charge noise. The dissipative dynamics of the system is numerically obtained from the master equation including the driving field and the environment. The latter is described by a bath of harmonic oscillators with ohmic spectrum at zero temperature \cite{Pekola2010,Solinas2010}. To be able to detect the effect of coherence loss, the pumping period must be smaller than the expected decoherence time. By choosing a superconducting island with high charging energy $E_C/k_B= 5$~K, this condition is fulfilled at frequencies as low as $0.5$~GHz, which is still in the regime of validity of the master equation \cite{Pekola2010,coupling_note}.

\begin{figure}
   \begin{center}
     \includegraphics{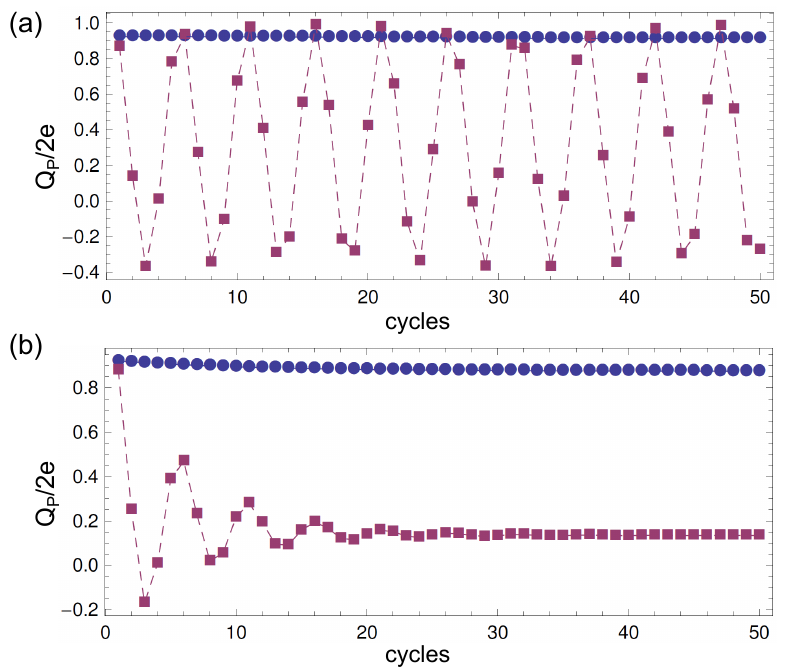}
   \end{center}
    \caption{(Color online) Pumped charge per cycle versus time for a case of constructive (squares) and destructive interference (circles), without decoherence (top) and with decoherence induced by zero-temperature gate-charge noise (bottom). Dashed lines are guides for the eye. The pumping frequency is 500 MHz for both cases, the phase biases are $\phi_c=2.4$ and $\phi_d=-1.45$, respectively. We have also assumed a residual Josephson coupling
$J_{min}=\min \{J_L,J_R\} =0.03J_{max}$.}
    \label{fig:decoherence}
\end{figure}

In Fig.~\ref{fig:decoherence}~(a) we present the expected pumped charge in the absence of noise over $50$ consecutive pumping cycles for constructive (squares) and destructive (circles) interference. The results are readily interpreted after Fig.~\ref{fig:2}~(b): When interference is destructive, the system starts every cycle in the ground state, so that the pumped charge is constant. On the contrary, constructive interference allows the system to be in different superpositions of the ground and the excited state, and this is reflected in an oscillating pumped charge from cycle to cycle.

When we include the environmental effects as in Fig.~\ref{fig:decoherence}(b), the behavior of the pumped charge in the constructive case changes dramatically, as oscillations are quickly damped due to the loss of coherence. At the same time, the pumped charge in the destructive case is only slightly affected by the environment. This is a direct result of the fact that the system stays mainly in the ground state, which was shown to be robust against decoherence \cite{Pekola2010,Solinas2010}. This suggests that the detrimental effects of decoherence can partly be overcome by a suitable choice of the phase bias. Most importantly, we notice that the large difference in the pumped charge makes it possible to clearly distinguish between the two cases.

In conclusion, we have proposed new type of Landau-Zener interferometry, based on geometric phases. Specifically, we have demonstrated this technique in a superconducting charge pump. For that case, we have shown that the geometric effects are controlled by the superconducting phase bias across the pump. They can be detected by measuring the pumped charge, and should persist in the presence of realistic decoherence.

We thank R. Fazio, M. M\"ott\"onen, S. Pugnetti, J. Salmilehto, M. Sill\"anp\"a\"a, and Y. Yoon for useful discussions. This work was supported by European Community's Seventh Framework Programme under Grant Agreement No.~238345 (GEOMDISS). P. S. acknowledges the Academy of Finland and Emil Aaltonen Foundation for financial support.


\begin{thebibliography}{29}
\bibitem [{Qp_note()}]{Qp_note}%
  \BibitemOpen
  \bibinfo {note} {For our case, the pumped charge can be safely identifed with
  the total charge transferred over a period.}\BibitemShut {Stop}%
\end{thebibliography}

%

\end{document}